\documentclass[11pt]{article}

\usepackage{graphics}
\usepackage{amsfonts}
\usepackage{amssymb}
\usepackage{amsmath}
\newcommand{\bo}[1]{\mathfrak{L}(#1)}
\newcommand{\bop}[1]{\mathfrak{L}^+(#1)}

\newcommand{\conj}[1]{{#1}^{\#}}
\newcommand{\tr}[1]{{\rm tr }\left(#1\right)}   %#1: argument of the trace
\newcommand{\flind}[1]{\mathfrak{L}\kern-8pt{-}(#1)}
\newcommand{\preflind}[1]{\mathfrak{L}_*\kern-12pt{-}\;\;(#1)}
\newcommand{\1}{\mathbf{1}}

\newcommand{\C}[1]{{\mathbb C}^{#1}}

\newcommand{\adj}[1]{{#1}^*}

\newtheorem{thm}{Theorem}

\newtheorem{cor}{Corollary}
\newtheorem{ex}{Example}

\newenvironment{pf}{\noindent{\it Proof.} }{\;\fbox{}\\}

\numberwithin{equation}{section}          

\title{On the structure of generators for non-Markovian Master Equations}
\author{Andrzej Kossakowski\\{\footnotesize\it Institute of Physics}\\{\footnotesize\it Nicholaus Copernicus University}\\{\footnotesize\it 87-100 Torun, Poland}\\{\footnotesize\it kossak@fizyka.umk.pl}\\ and \\Rolando Rebolledo \footnote{Research supported by the Bicentennial Program PBCT-ACT13, Chile and the Program of Visiting Scholars of the Pontificia Universidad Cat\'olica de Chile}\\{\footnotesize\it Laboratorio de An\'alisis Estoc\'atico}\\{\footnotesize\it Facultad de
Matem\'aticas}\\{\footnotesize\it Pontificia Universidad Cat\'olica de Chile}\\{\footnotesize\it Casilla 306, Santiago 22,
Chile}\\{\footnotesize\it rrebolle@puc.cl}}
\date{}
\begin{document}
\maketitle
\begin{abstract}
 Complete characterization of complete positivity preserving non-Markovian master equations is presented. \end{abstract}
\section{Introduction}
The study of time evolution of quantum open systems plays an important role in quantum information. The interaction between the system and its environment leads to phenomena of decoherence and dissipation \cite{Alicki:1987yo,Breuer:2002uq,Nielsen:2000ly}.

The Nakajima-Zwanzig projector operator method \cite{Nakajima:1958fx,Zwanzig:1960zh} makes possible to derive an exact equation for the reduced density matrix from the von Neumann equation of the composed system. The resulting non-Markovian master equation is mostly of formal interest since its solution cannot be written down explicitely, in closed form. In contrast, when the Markovian approximation is used, that is, when memory effects are neglected, the resulting Markovian master equation \cite{gorkossud,lindblad76} has a simpler form and complete positivity is preserved during the evolution \cite{Kraus:1983ta}.

A challenge for the non-Markovian theory of open quantum systems is to obtain a characterization of a class of evolutions which preserves complete positivity and captures reservoir memory effects at the same time.

A variety of non-Markovian master equations have been proposed (cf. \cite{Breuer:2002uq,Shibata:1977mi,Imamoglu:1994pi,Royer:1996ff,Royer:2003lh,Barnett:2001fu,Breuer:2001qy,Breuer:2001kh,Breuer:1999rq,Breuer:2004fc,Breuer:2004ss,Breuer:2006gf,Maniscalco:2005ye,Maniscalco:cq,Wilkie:2000qo,Wilkie:2001tw,Breuer:qc,Budini:2004jl,Budini:il,Budini:zt,Shabani:2005gb,Lee:2004mb,Pulo:fm,Kossakowski:2007eu}). However, the complete positivity of the resulting time evolution remains an important problem \cite{Kossakowski:2008eu,Breuer:fv}. In the present paper the structure of non-Markovian master equations preserving complete positivity is given. 
\section{Notations and preliminaries}
To avoid technical difficulties connected with infinite dimensions, we restrict ourselves to the $d$-dimensional Hilbert space $\C{d}$ of complex vectors with the scalar product $\langle\cdot,\cdot\rangle$ and elements denoted $e,x,y,z,\ldots$. The extension to infinite dimensional Hilbert spaces is currently being investigated and will appear in a forthcoming paper.

The $C^*$--algebra of linear operators on $\C{d}$ will be denoted $M_{d}=M_{d}(\C{})$, or $\bo{\C{d}})$. Elements of $M_{d}$ will be denoted by $a,b,c,\ldots$ and its unit by $\1$. $M_{d}$ is also a Hilbert space under the scalar product $\langle a,b\rangle=\tr{\adj{a}b}$.

The $C^*$--algebra of linear maps from $M_{d}$ into $M_{d}$ will be denoted by $\bo{M_{d}}$, its elements are denoted by capital letters $A,B,C,\ldots$ and the identity map in $\bo{M_{d}}$ by $I$. The conjugation (duality) $\conj{\cdot}$ in $\bo{M_{d}}$ is defined by the relation:
\begin{equation}
\langle \conj{A}a,b\rangle=\langle a,Ab\rangle,
\end{equation}
for all $a,b\in M_{d}$.

This operation endows the following property: the relations

\begin{equation}
A\1=\1,\;\;L\1=0,
\end{equation}
and 
\begin{equation}
\tr{\conj{A}a}=\tr{a},\;\;\tr{\conj{L}a}=0,
\end{equation}
are equivalent.

The cone of all completely positive maps on $M_{d}$ will be denoted by $\bop{M_{d}}$. Within this paper we make an intensive use of Laplace Transform Theory. In general we reserve the symbol $\widehat{\cdot}$ and the variable $p$ for Laplace transforms. In general we will consider scalar functions $f(t)$ \emph{positive} and integrable on $[0,\infty]$ and denote its Laplace transform by $\widehat{f}(p)=\int_{0}^\infty\;dt\exp (-pt)f(t)$. On the other hand, measurable $\bo{M_{d}}$--valued functions will be written like $A_{t}\in\bo{M_{d}}$, $t\geq 0$. For such a function the Laplace transform exists and will be denoted $\hat{A}_{p}$.

\section{Non-Markovian master equations}
The reduced dynamics can be studied equivalently in the Schr\"odinger or the Heinsenberg pictures. Suppose that $A_{t}:M_{d}\to M_{d}$ describes the reduced dynamics in the Heisenberg picture, then it should satisfy the following conditions:
$A_{t}\in\bop{M_{d}}$, $A_{t}\1=\1$, for all $t\geq 0$, and $A_{0}=\lim_{t\to 0}A_{t}=I$. In the Schr\"odinger picture these relations are given in terms of $\conj{A}_{t}$, $t\geq 0$.

In the present paper the reduced dynamics is investigated under the assumption that $A_{t}$ is the solution of a non-Markovian master equation of the form:
\begin{equation}\label{nonmarkovian}
\frac{dA_{t}}{dt}=LA_{t}+\int_{0}^tdsL_{t-s}A_{s},
\end{equation}
with the initial condition $A_{0}=I$. 

The normalization condition $A_{t}\1=\1$ implies the equality
\begin{equation}
L_{t}\1=0\,,\qquad L\1=0\,.
\end{equation}
Let us observe that $L$ can be formally absorbed in $L_t$ by the transformation
\begin{equation}
 L_t\;\longmapsto\;L_t'=L_t+2\delta(t)\,L\,.
\end{equation}
In the present paper the equation
\begin{equation}\label{kr3.4}
\frac{d A_{t}}{dt}=\int_{0}^tds L_{t-s}A_{s}.
\end{equation}
will be considered. 

The map $L_{t}$ will be referred as the \emph{generator of the Master Equation}. One of the fundamental problems of non-Markovian master equations is to find conditions on $L_{t}$ that ensure that the time evolution resulting from \eqref{kr3.4} is completely positive. The result of our previous paper \cite{Kossakowski:2008eu} can be reformulated in the following manner:
\begin{thm}
Let be given a family of maps $Z_{t}\in\bo{M_{d}}$, $(t\geq 0)$, such that 
\begin{equation}\label{3.5}
L_{t}=B_{t}-Z_{t},
\end{equation}
where $B_t\in \bop{M_{d}}$ for all $t\geq 0$, and
\begin{equation}\label{3.6}
L_{t}\1=B_{t}\1-Z_{t}\1=0.
\end{equation}
Then the time evolution $A_{t}$ resulting from \eqref{kr3.4} is completely positive if the solution of the normalization equation
\begin{equation}\label{kr3.7}
\frac{d}{dt}N_{t}=-\int_{0}^tdsZ_{t-s}N_{s};\;N_{0}=I,
\end{equation}
is completely positive.
\end{thm}

This version of the theorem leads to the difficult question of obtaining conditions on $Z_{t}$ guaranteeing that $N_{t}$ is completely positive for any $t\geq 0$.  The case $Z_{t}=\frac{1}{2}(c_{t}a+a\adj{c_{t}})$ and $c_{t}\adj{c}_{t}=\adj{c}_{t}c_{t}$ has been investigated in \cite{Breuer:fv}. The construction of $Z_{t}$ for which $N_{t}$ is completely positive is given in the following Theorem.
\begin{thm}
Suppose the solution $N_{t}$ of the normalization equation is completely positive and given in the following form 
\begin{equation}\label{ncanonical}
N_{t}=I-\int_{0}^tdsF_{s},
\end{equation}
where $t\mapsto F_{t}$ is an integrable $\bo{M_{d}}$-valued function.

Then $Z_{t}$ is the solution of the integral equation
\begin{equation}\label{3.8}
\int_{0}^tdsN_{t-s}Z_{s}=F_{t}.
\end{equation}

\end{thm}

\begin{pf}
Taking the Laplace transform of \eqref{kr3.7} yields
\begin{equation}\label{3.10}
\widehat{N}_{p}=(pI+\widehat{Z}_{p})^{-1}
\end{equation}
By hypothesis
\begin{equation}\label{3.12}
\widehat{N}_{p}=\frac{1}{p}(I-\widehat{F}_{p}),
\end{equation}
so that comparing with \eqref{3.10} we get
\begin{equation}\label{3.11}
\widehat{Z}_{p}=p\widehat{F}_{p}(I-\widehat{F}_{p})^{-1}.
\end{equation}
It is a straightforward computation to verify that the map $Z_{t}$ is the solution of \eqref{3.8}.
\end{pf}
The Theorem 1 has been recently generalized by Breuer and Vaccini \cite{Bre09}.
\begin{thm}[Breuer-Vaccini]
 Suppose that the generator $L_t$ is given in the form
\begin{equation}\label{ko3.13}
 L_t=B_t-Z_t\,,\qquad t\geq 0\,,
\end{equation}
where 
\begin{equation}\label{ko3.14}
 L_t\1=B_t\1-Z_t\1=0\,,
\end{equation}
the solution $A_t$, $t\geq0$, of (\ref{kr3.4}) is completely positive if the solution $N_t$, $t\geq0$, of the normalization equations (\ref{kr3.7}) is complete positive and the map
\begin{equation}\label{ko3.15} 
 \int\limits_0^t ds N_{t-s}B_s
\end{equation}
is completely positive for all $t\geq 0$.
\end{thm}
The application of Theorems 2 and 3 is illustrated as follows.
\begin{ex}{\em
Suppose that $N_{t}$ is of the form 
\begin{equation}\label{3.15}
N_{t}=\left(1-\int_{0}^tdsf(s)\right)I,
\end{equation}
where $f(s)$ is a positive measurable scalar function, such that
\begin{equation}\label{3.16}
\int_{0}^\infty ds f(s)\leq 1,
\end{equation}
this means that one has $F_{t}=f(t)I$. It follows from \eqref{3.8} that
\begin{equation}
Z_{t}=\kappa (t) I,
\end{equation}
where $\kappa(t)$ is given in terms of its Laplace transform 
\begin{equation}\label{3.18}
\widehat{\kappa}(p)=\frac{p\widehat{f}(p)}{1-\widehat{f}(p)}
\end{equation}
Let $B_t$, $t\geq0$, be given in the form 
\begin{equation}\label{ko3.20}
 B_t=\kappa(t)B\,,
\end{equation}
where $B$ is completely positive and normalized ($B1=1$) map on $M_d$. Using (\ref{3.15}), (\ref{3.18})
and (\ref{ko3.20}) one finds 
\begin{equation}\label{ko3.21}
 \int\limits_0^t ds N_{t-s}B_s=f(t)B\,,
\end{equation}
i.e., the condition (\ref{ko3.15}) is satisfied.
}
\end{ex}
\begin{cor}
The family $L_{t}$, $t\geq 0$, of maps:
\begin{equation}\label{3.19}
L_{t}=\kappa (t)(B-I),
\end{equation}
where $B$ is completely positive and normalized map on $M_d$,
is the generator of a non-Markovian master equation provided $\kappa (t)$ is the solution of 
the integral equation
\begin{equation}
\int\limits_{0}^tds g(t-s)\kappa (s)=f(t),
\end{equation}
where
\begin{equation}
g(t)=1-\int\limits_{0}^tdsf(s),
\end{equation}
and $f(t)$ is a positive measurable function satisfying \eqref{3.16}.
\end{cor}
The above result can be easily generalized. Let $B$ be as above, and $F_t$ defined by (\ref{ncanonical}) is
completely positive for all $t\geq 0$, then
\begin{equation}
 {\cal L}_t=Z_t(B-I)\,,
\end{equation}
where $Z_t$ is given by (\ref{3.8}), is the generator of non-Markovian master equation.

The reduced dynamics $A_t$, $t\geq 0$ is characterized as a family of completely positive and normalized maps which is the solution of  (\ref{kr3.4}) under the initial condition $A_0=I$. The reduced dynamics can also be characterized as follows
\begin{thm}
  Let $A_t$, $t\geq0$ be the reduced dynamics, then $A_t$ has the representation
\begin{equation}\label{kr3.35a}
 A_t\;=\;I+\int\limits_0^tds G_s\,,
\end{equation}
where 
\begin{equation}
 G_s(\1)=0\,.
\end{equation}
\end{thm}
{\it Proof.}\quad The reduced dynamics is defined as follows
\begin{equation}\label{kr3.37a}
 A_t a\;=\;{\rm tr}_{\cal H}\omega[e^{tL}(a\otimes \1_{\cal H})]\,,
\end{equation}
where $e^{tL}$ is completely positive and normalized semigroup on $M_d\otimes B({\cal H})$, $\omega$ is a fixed
normal state on $B({\cal H})$ and $\1_{\cal H}$ is the identity map on ${\cal H}$.
It follows from (\ref{kr3.37a}) that (\ref{kr3.35a}) holds with
\begin{equation}
 G_ta\;=\;{\rm tr}_{\cal H}\omega[Le^{tL}(a\otimes\1_{\cal H})]\,.
\end{equation}
\begin{thm}
 Let $A_t$, $t\geq0$ be a family of completely positive and normalized maps satisfying the equation
\begin{equation}\label{kr3.34}
 \frac{dA_t}{dt}\;=\;\int\limits_0^t ds L_{t-s}A_s\,,\qquad A_0=I
\end{equation}
i.e.,
\begin{equation}\label{kr3.35}
 \widehat{A}_p\;=\;\frac{1}{\1-\widehat{L}_p}\,,
\end{equation}
then the relations
\begin{equation}\label{kr3.36}
 A_t\;=\;\1+\int\limits_0^t ds G_s
\end{equation}
where
\begin{equation}\label{kr3.37}
 G_s(1)\;=\;0\,,
\end{equation}
and
\begin{equation}\label{kr3.38}
 \widehat{L}_p\;=\;\frac{p\widehat{G}_p}{\1+\widehat{G}_p}
\end{equation}
are equivalent.
\end{thm}
{\it Proof.} From (\ref{kr3.36}) ome obtains
\begin{equation}\label{kr3.39}
 \widehat{A}_p\;=\;\frac1{p}(\1+\widehat{G}_p)\,.
\end{equation}
From (\ref{kr3.35}) and (\ref{kr3.39}) it follows that relation (\ref{kr3.38}) hold. On the other hand inserting (\ref{kr3.38}) into (\ref{kr3.35}) one finds (\ref{kr3.36}).

\begin{ex}\rm 
 Let $A_t=e^{tL}$, $t\geq0$ be a completely positive semigroup. One can write $A_t$ in the form
\begin{equation}\label{kr3.40}
 A_t\;=\;\1+\int\limits_0^t ds G_s
\end{equation}
where
\begin{equation}\label{kr3.41}
 G_t\;=\;L\,e^{tL}\,.
\end{equation}
Using the relation (\ref{kr3.38}) and (\ref{kr3.41}) one finds
\begin{equation}\label{3.42}
 \widehat{L}_p=L\,,
\end{equation}
i.e.
\begin{equation}\label{kr3.43}
 L_t\;=\;2\delta(t)L
\end{equation}
and (\ref{kr3.34}) takes the form
\begin{equation}\label{kr3.44}
 \frac{dA_t}{dt}\;=\;LA_t\,.
\end{equation}
\end{ex}
\begin{ex}\rm
Let $L_1,\ldots,L_n$ be generators of completely positive semi-groups and $x_1,\ldots,x_n\geq0$,
\begin{equation}\label{kr3.45}
 \sum_{j=1}^n x_j\;=\;1\,.
\end{equation}
 Let $A_t$, $t\geq 0$ be the family of completely positive normalized maps defined as follows
\begin{equation}\label{kr3.46}
 A_t\;=\;\sum_{j=1}^n x_j\,e^{tL_j}\,.
\end{equation}
$A_t$ can be rewritten in the form
\begin{equation}\label{kr3.47}
 A_t\;=\;\1+\int\limits_0^t ds G_s\,,
\end{equation}
where
\begin{equation}\label{kr3.48}
 G_t\;=\;\sum_{j=1}^nx_jL_j\,e^{tL_j}\,.
\end{equation}
Taking the Laplace transform of $G_t$ and using (\ref{kr3.38}) one finds
\begin{equation}\label{kr3.49}
 \widehat{L}_p\;=\;\frac{p\sum_{j=1}^n x_jL_j(p-L_j)^{-1}}{\1+\sum_{j=1}^n x_j(p-L_j)^{-1}}\,.
\end{equation}
Suppose that $[L_i,L_j]=0$, $i,j=1,\ldots,n$ and consider the special cases $n=2$ and $n=3$.

For $n=2$ it follows from (\ref{kr3.49}) that
\begin{equation}\label{kr3.50}
 \widehat{L}_p\;=\;x_1L_1+x_2L_2+\frac{x_1x_2(L_1-L_2)^2}{p-(x_1L_2+x_2L_1)}\,.
\end{equation}
In the case $n=3$ one obtains
\begin{equation}\label{kr3.51}
\widehat{L}_p\;=\;L+\frac{\widehat{B}_p}{\widehat{C}_{p}}\,,
\end{equation}
where
\begin{eqnarray}
L &=& x_1L_1+x_2L_2+x_3L_3\,,\\
\widehat{B}_p &=& p(x_1L_1^2+x_2L_2^2+x_3L_3^2-L^2)+L_1L_2L_3 \nonumber\\
&&-\;L(x_1L_2L_3+x_2L_1L_3+x_3L_1L_2)\,,\\
\widehat{C}_p &=& p^2-p(L_1+L_2+L_3-L)+x_1L_2L_3+x_2L_1L_3+x_3L_1L_2\,.\qquad\mbox{}
\end{eqnarray}
\end{ex}
\begin{ex}\rm
Let $L_1,\ldots,L_n$ be generators of completely positive semi-groups and $x_1(t),\ldots,x_n(t)$ be nonnegative functions such that
\begin{equation}\label{kr3.55}
 \sum_{j=1}^n \int\limits_0^t ds x_j(s)\;\leq\;1\,.
\end{equation}
 Let us consider the family $A_t$, $t\geq 0$  of completely positive and normalized maps defined as follows
\begin{equation}\label{kr3.56}
 A_t\;=\;\1\Big(1-\sum_{j=1}^n\int\limits_0^t ds x_j(s)\Big)+\sum_{j=1}^n\int\limits_0^t ds x_j(s)\,e^{sL_j}\,.
\end{equation}
The formula (\ref{kr3.56}) can be rewritten in the form
\begin{equation}\label{kr3.57}
 A_t\;=\;\1+\int\limits_0^t ds G_s\,,
\end{equation}
where
\begin{equation}\label{kr3.58}
 G_t\;=\;\sum_{j=1}^nx_j(t)(e^{tL_j}-\1)\,.
\end{equation}
The generator $\widehat{L}_p$ is given by (\ref{kr3.38}) and it cannot be written explicitely.
\end{ex}
The examples 3, 4 and 5 show that, in general, it is rather difficult to write down the generator $L_t$ explicitely. However, if the family $A_t$, $t\geq0$ of completely positive maps can be represented in the form (\ref{kr3.36}) then $A_t$ is the solution of non-Markovian master equation (\ref{kr3.34}).

\section*{Acknowledgements}
A part of this work was presented at the Torun 40th Symposium on Mathematical Physics. This symposium considered a special session in honor of Professor Andrzej Kossakowski. R.R. wish to acknowledge the kind invitation and hospitality of the Organizing Committee of the Symposium, which allowed him to express his deep gratitude to Andrzej for all the time afforded to a very fruitful joint scientific collaboration and friendship.

A.\,K. acknowledges the warm hospitality during the stay in PUC and a partial support by the Polish
Ministry of Science and Higher Education Grant No 3004/B/H03/2007/33.

\def\cprime{$'$} \def\cprime{$'$}


\begin{thebibliography}{10}

\bibitem{Alicki:1987yo}
Alicki, R.; Lendi, K. {\em Quantum Dynamical Semigroups and Applications}, Vol.
   286 of {\em Lecture Notes in Phys.;}
\newblock Springer-Verlag, 1987.

\bibitem{Breuer:2002uq}
Breuer, H.-P.; Petruccione, F. {\em The theory of open quantum systems;}
\newblock Oxford University Press: New York, 2002.

\bibitem{Nielsen:2000ly}
Nielsen, M.~A.; Chuang, I.~L. {\em Quantum computation and quantum
  information;}
\newblock Cambridge University Press: Cambridge, 2000.

\bibitem{Nakajima:1958fx}
Nakajima, S. {\em Prog. Theor. Phys.} {\bf 1958}, {\em 20}, 948.

\bibitem{Zwanzig:1960zh}
Zwanzig, R. {\em J. Chem. Phys.} {\bf 1960}, {\em 33}, 1338.

\bibitem{gorkossud}
Gorini, V.; Kossakowski, A.; Sudarshan, E. {\em J. Math. Phys.} {\bf 1976},
  {\em 17}, 821--825.

\bibitem{lindblad76}
Lindblad, G. {\em Commun. Math. Phys.} {\bf 1976}, {\em 48}, 119--130.

\bibitem{Kraus:1983ta}
Kraus, K. {\em States, Effects and Operations, Fundamental notions of Quantum
  Theory;}
\newblock Academics, Berlin, 1983.

\bibitem{Shibata:1977mi}
Shibata, N.; Takahashi, Y. {\em J.Stat.Phys.} {\bf 1977}, {\em 17}, 171.

\bibitem{Imamoglu:1994pi}
Imamoglu, A. {\em Phys. Rev. A} {\bf 1994}, {\em 50}, 3650.

\bibitem{Royer:1996ff}
Royer, A. {\em Phys. Rev. Lett.} {\bf 1996}, {\em 77}, 3272.

\bibitem{Royer:2003lh}
Royer, A. {\em Phys. Lett. A} {\bf 2003}, {\em 315}, 335.

\bibitem{Barnett:2001fu}
Barnett, M.; Stenholm, S. {\em Phys. Rev. A} {\bf 2001}, {\em 64}, 033808.

\bibitem{Breuer:2001qy}
Breuer, H.-P.; Kappler, B.; Petruccione, F. {\em Ann. Physics} {\bf 2001}, {\em
  291}(1), 36--70.

\bibitem{Breuer:2001kh}
Breuer, H.~P.; Petruccione, F. {\em Phys. Rev. A} {\bf 2001}, {\em 63}, 032102.

\bibitem{Breuer:1999rq}
Breuer, H. P.;~Kappler, B.; Petruccione, F. {\em Phys. Rev. A} {\bf 1999}, {\em
  59}, 1633.

\bibitem{Breuer:2004fc}
Breuer, H.~P. {\em Phys. Rev. A} {\bf 2004}, {\em 69}, 022115.

\bibitem{Breuer:2004ss}
Breuer, H.~P. {\em Phys. Rev. A} {\bf 2004}, {\em 70}, 012106.

\bibitem{Breuer:2006gf}
Breuer, H.-P.; Gemmer, J.; Michel, M. {\em Phys. Rev. E (3)} {\bf 2006}, {\em
  73}(1), 016139, 13.

\bibitem{Maniscalco:2005ye}
Maniscalco, S. {\em Phys. Rev. A} {\bf 2005}, {\em 72}, 024103.

\bibitem{Maniscalco:cq}
Maniscalco, S.; Petruccione, F. {\em Phys. Rev. A} {\bf 2006}, {\em 73},
  012111.

\bibitem{Wilkie:2000qo}
Wilkie, J. {\em Phys. Rev. E} {\bf 2000}, {\em 62}, 8808.

\bibitem{Wilkie:2001tw}
Wilkie, J. {\em J. Phys. A} {\bf 2001}, {\em 114}, 7736.

\bibitem{Breuer:qc}
Breuer, H.~P. {\em Phys. Rev. A} {\bf 2007}, {\em 75}, 022103.

\bibitem{Budini:2004jl}
Budini, A. {\em Phys. Rev. A} {\bf 2004}, {\em 69}, 042107.

\bibitem{Budini:il}
Budini, A. {\em Phys. Rev. A} {\bf 2006}, {\em 74}, 053815.

\bibitem{Budini:zt}
Budini, A. quant-ph/0611222.

\bibitem{Shabani:2005gb}
Shabani, A.;~Lidar, D. {\em Phys. Rev. A} {\bf 2005}, {\em 71}, 020101.

\bibitem{Lee:2004mb}
Lee, J.;~Kim, I. A. D. M. A.~H.; Kim, M. {\em Phys. Rev. A} {\bf 2004}, {\em
  70}, 024301.

\bibitem{Pulo:fm}
P{\"u}lo, J.;~Maniscalco, S.~K. quant-ph/07064438, {\bf 2007}.

\bibitem{Kossakowski:2007eu}
Kossakowski, A.; Rebolledo, R. {\em Open Syst. Inf. Dyn.} {\bf 2007}, {\em
  14}(3), 265--274.

\bibitem{Kossakowski:2008eu}
Kossakowski, A.; Rebolledo, R. {\em Open Syst. Inf. Dyn.} {\bf 2008}, {\em
  15}(2), 135--141.

\bibitem{Breuer:fv}
Breuer, H.-P; Vaccini, B. Phys. Rev. Lett {\bf 101}, 140402 (2008).
\bibitem{Bre09}
Breuer H.-P., Vaccini, B., arXiv: quant-ph 09022318.

\end{thebibliography}
\end{document}